\documentclass[twocolumn,superscriptaddress,showpacs,amsmath,amssymb,pra,longbibliography]{revtex4-1}

\usepackage[pdftex]{graphicx} 
\usepackage{dcolumn}  
\usepackage{bm}       
\usepackage[usenames,dvipsnames]{color}
\definecolor{URLCOL}{rgb}{0,0.52,0.83} 
\definecolor{LINKCOL}{rgb}{0.05,0.5,0} 
\definecolor{orange}{rgb}{0.6,0.3,0} 
\definecolor{CITECOL}{rgb}{0.25,0,0.48} 
\usepackage{epstopdf}
\usepackage{amsmath}

\usepackage[pdftex,bookmarks,breaklinks,bookmarksopen,bookmarksnumbered,colorlinks,linkcolor=LINKCOL,linktocpage,citecolor=CITECOL,urlcolor=URLCOL,pdfpagemode=UseOutline,pdftex]{hyperref}

\usepackage{fancyhdr}
\fancypagestyle{titlepage}
{\chead{file: \jobname}\rhead{Total pages: \pageref{page:end}}}

\usepackage{booktabs}
\usepackage{natbib}

\definecolor{TITLECOL}{rgb}{0.1,0.2,0.7} 
\definecolor{SECOL}{rgb}{0.1,0.2,0.7} 
\definecolor{CONTENTSCOL}{rgb}{0.1,0.2,0.7} 
\definecolor{SSECOL}{rgb}{0.25,0,0.48} 
\definecolor{SSSECOL}{rgb}{0.2,0.08,0.53} 
\definecolor{FINCOL}{rgb}{0.01,0.3,0.07} 

\def\coloredtitle#1{\title{\textcolor{TITLECOL}{#1}}} 
\def\coloredauthor#1{\author{\textcolor{CITECOL}{#1}}} 

\definecolor{URLCOL}{rgb}{0,0.17,0.43} 
\definecolor{LINKCOL}{rgb}{0.05,0.4,0} 
\definecolor{CITECOL}{rgb}{0.35,0,0.48} 

\def\bea{\begin{eqnarray}}
\def\eea{\end{eqnarray}}
\def\ben{\begin{equation}}
\def\een{\end{equation}}
\def\benu{\begin{enumerate}}
\def\enu{\end{enumerate}}

\def\bei{\begin{itemize}}
\def\eei{\end{itemize}}
\def\beit{\begin{itemize}}
\def\eit{\end{itemize}}
\def\benu{\begin{enumerate}}
\def\enu{\end{enumerate}}

\def\half{\frac{1}{2}}

\def\tau{T}

\def\sec#1{\section{\textcolor{SECOL}{#1}}}

\begin{document}

\coloredtitle{Confirmation of the PPLB derivative discontinuity: \\
Exact chemical potential at finite temperatures of a model system}
\coloredauthor{Francisca Sagredo}
\affiliation{Department of Chemistry, University of California, Irvine, CA 92697}
\date{\today}

\coloredauthor{Kieron Burke}
\affiliation{Department of Chemistry, University of California, Irvine, CA 92697}
\affiliation{Department of Physics and Astronomy, University of California, Irvine, CA 92697}
\date{\today}
\begin{abstract}
The landmark 1982 paper of Perdew, Parr, Levy, and Balduz (often called PPLB) laid the
foundation for our modern understanding of the role of the derivative discontinuity in density
functional theory, which drives much development to account for its effects.
A simple model for the chemical potential at vanishing temperature played a crucial role
in their argument.  We investigate the validity of this model in the simplest
non-trivial system to which it can be applied and which can be easily solved exactly, the Hubbard dimer.
We find exact agreement in the crucial zero-temperature limit, and show the model remains accurate
for a significant range of temperatures.  We identify how this range depends on the strength of
correlations. We extend the model to approximate free energies accounting for the derivative
discontinuity, a feature missing in standard semilocal approximations. We provide a correction to this approximation to yield even more accurate free energies. We discuss the
relevance of these results for warm dense matter.
\end{abstract}

\maketitle

A crucial concern for density functional theory (DFT) calculations of semiconductor solids in the 1980's was
whether the systematic underestimate of the band gap represented a limitation
of approximations, or a fundamental deficiency of Kohn-Sham (KS) DFT \cite{KS65}.
The paper of Perdew, Parr, Levy, and Balduz (PPLB) \cite{PPLB82} argued clearly that the band gap of
a pure KS DFT calculation does not in general match the fundamental gap, even if
the exact functional is used \cite{WSBW13}. In the decades since, this
understanding has become a cornerstone of modern DFT. Its generalization to 
include spin-degrees of freedom \cite{AY06,Pb86} has led to approximate
functionals that explicitly account for delocalization errors \cite{SCY18,M05,PCVJ92,PY84,HKSG17}.  A deep but more accessible background article was written by
Perdew only a few years later \cite{Pb85}, and later with Parr and Yang \cite{PY89}. 

A vital step in the logic of this work is the introduction of the grand canonical (gc)
ensemble to couple the electronic system of interest to a thermodynamic bath. At any
finite temperature \cite{M65}, even the exact functional is a smooth continuous function of the average
particle number $\mathcal{N}$, but develops steps at integers that sharpen as the temperature
is lowered.  As the temperature tends to zero, the gc ensemble reduces
to a linear ensemble between the integers, which in turn leads to the modern theory of ground-state DFT
for non-integer particle numbers \cite{DG84}.  As bonds are stretched, or electrons added and removed
from solids, the energetic consequences can be directly related to the discontinuities
in the slope of the energy as a function of $\mathcal{N}$.  Many of these effects, like charge transfer in molecular systems are missed
by semi-local approximations which, by construction, have no discontinuities \cite{PS84,T03,M05,TFSB05,KBE06,KBY07,HK12,NICP13}. Hence 
the ongoing desire to create approximations that can quantitatively account for such effects \cite{SCY18}.

In the last two decades, DFT calculations at finite temperature have helped revolutionize
the field of warm dense matter, by producing chemically specific quantitative predictions for present-day
shock experiments \cite{SSB16}. Their legitimacy stems from Mermin's theorem \cite{M65}, which generalizes the
Hohenberg-Kohn theorem \cite{HK64} to non-zero temperature, and therefore non-integer average particle numbers. Modern warm dense calculations run standard solid-state
codes to solve the KS equations, with finite temperature Fermi occupations. These are used to model shock experiments, \cite{RMCHM10}, understand planetary cores \cite{D03,NHKF08}, and even model inertial confinement fusion \cite{KDD08,KHBH01}.  Thus
there is rapidly growing interest in the theory of equilibrium electronic structure beyond the ground-state.

\begin{figure}[htb]
  \includegraphics[width=\linewidth]{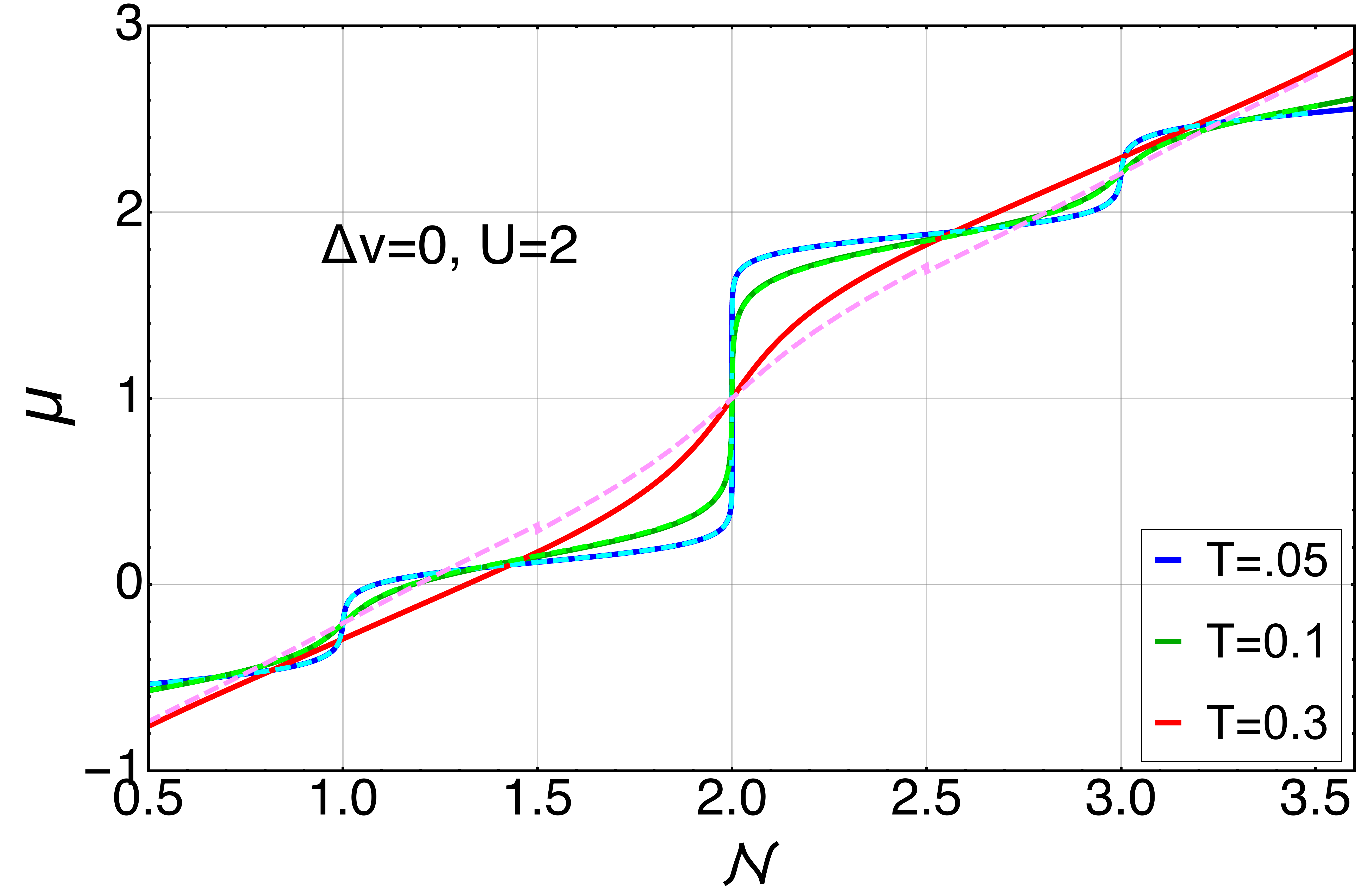}
  \caption{Exact (solid) and PPLB (dashed) chemical potentials in the symmetric ($\Delta v = 0$) Hubbard dimer at $U=2$ with various temperatures $\tau$, with $2t=1$.}
  \label{mu}
\end{figure}

In the current work, we calculate the chemical potential of a simple model system
exactly, as a function of average particle number and temperature. We confirm the
ansatz behind the PPLB work:  their approximation to the chemical potential becomes
relatively exact for all particle numbers as the temperature $\tau \rightarrow 0$. For our simple system, we also explore up to what temperatures the PPLB formula works. We also generalize the PPLB model to extract the free energy, and explore its accuracy.
We explore how the strength of correlations affect the accuracy of the PPLB approximation. Finally, we give a generalization that corrects an
obvious limitation of the PPLB model.

Fig. \ref{mu} illustrates our key results nicely. The chemical potential is very smooth at higher temperatures,
but steps develop around integer particle numbers as the temperature is lowered. The PPLB approximation becomes exact in
the limit of zero temperature, matching the exact derivative discontinuities, but in this system, the PPLB model continues to work well even at significant warm temperatures, as shown for $\tau=0.3$. Ironically, unlike local and semi local DFT approximations at zero temperature, the PPLB model's largest error is
at half-integers, where it incorrectly jumps discontinuously. The error at these half integer particle numbers is due to its dependence on the nearest integer to $\mathcal{N}$.

We begin with a brief recap of the PPLB argument.  For a finite system in contact with a bath at temperature $\tau$,
with which it can exchange both electrons and energy, its equilibrium properties are given by the gc
ensemble.  The gc partition function sums over all particle numbers, $N$, and eigenvalues \cite{Schwabl02}. For sufficiently low temperatures,
the ground-state energy will dominate over all others for each value of $N$, so all excited state contributions can be ignored.  Moreover, the convexity of $\mu N - E$
ensures that, for $\mathcal{N}=M+\nu$, where $|\nu| \leq \half$, the partition function will be dominated by only three contributions,
from $M-1$, $M$, and $M+1$. Including just these three terms, one can solve explicitly for the chemical potential, $\mu$, to find the PPLB approximation
\ben
\mu^{^{PPLB}}
=-\tau \log\bigg( \frac{- \nu+\sqrt{\nu^{2}+4h^{^{-}}h^{^{+}} (1-\nu^{2})}}{2h^{^{-}}(1+\nu)}\bigg), \\ 
\label{muPPLB}
\een
\noindent where
\ben
h^{\pm}= \frac{g_{_{M\pm 1}} \exp{(-(E_{M}-E_{M\pm1})/\tau)} }{g_{_{M}}},
\een
and $g_{_{M}}$ and $E_{M}$ are the degeneracy and ground state energy for $M$ particles. As stated in Ref.\cite{Pb85}, this form was derived only for the limit as $\tau \rightarrow 0$. However we will see that it can in fact be used for finite temperatures. 

The inclusion of degeneracies first appears in Ref. \cite{Pb85}.  In the zero-temperature limit, we make note of a few things. First, that $-\mu$ is simply the Mulliken electronegativity, $\chi$, and second, the iconic results of $\mu = -I$ below an integer and $-A$ above, where $I$ and $A$ are the ionization potential and electron affinity, respectively. Moreover, $\mu= -(I+A)/2$ at the integers. This determines the plateaus in Fig. \ref{mu}
since, at zero temperature, $\mu=\partial E/\partial \mathcal{N}$, so the size of the steps in $\mu$ are
the derivative discontinuities in $E(\mathcal{N})$.  In KS DFT, only part of these steps is in the KS kinetic energy,
leaving crucial contributions in the ubiquitous exchange-correlation (XC) energy.  As XC potentials are functional
derivatives of XC energies, they have spontaneous steps as the particle number moves across an integer \cite{HKSG17}, and sharp features in the middle of strongly stretched bonds \cite{HKSG17}.

It is difficult to imagine calculating the analog of Fig. \ref{muPPLB} sufficiently accurately from any first-principles
Hamiltonian, as it requires sums over all states and all particle numbers, including those in the
continuum.  But the two-site Hubbard model has a tiny Fock space, with only 16 states total.
Its Hamiltonian is
\ben
\hat{H} = -t\, \sum_{\sigma} \: (\hat{c}_{1\sigma}^{\dagger}\hat{c}_{2\sigma} + h.c) +
U \sum_{i} \hat{n}_{i,\uparrow\,}\hat{n}_{i,\downarrow} + \sum_i v_i \hat{n}_i  ,
\label{eq:HH}
\een

\noindent where $\hat{c}_{1\sigma}^{\dagger}$ and $\hat{c}_{2\sigma}$ are the creation and annihilation operators for each site,
$t$ is an electron hopping energy, $U$ is 
the repulsion between the particles in each site, and 
$\Delta v= v_{2}-v_{1}$ is the difference in external potential on the left and right sites \cite{CFSB15}.
We always choose $2t=1$ to set the energy scale. 
In chemistry,
the symmetric case ($\Delta v =0$) is the Hamiltonian for H$_2$ in a minimal basis. For any $\Delta v$, $U=0$ is the tight binding limit. Over the decades, the dimer has been used as a model for testing many
concepts in KS DFT \cite{AG02}. 
The density is characterized by a single number, $\Delta n= n_{1}-n_{2}$. 
The analyticity of this model system makes it perfect for testing fundamental aspects in DFT. Recently, it formed the basis of reviews of both ground-state DFT \cite{CFSB15} and linear response TDDFT \cite{CFMB18}. The dimer was used to check approximations in ensemble DFT \cite{SB18}, to illustrate several theorems in finite temperature DFT \cite{SPB16}, and even to study magnetism \cite{U18}.
Here, we use it simply as the simplest non-trivial model of interacting electrons to which we can apply
quantum statistical mechanics, and thus test the PPLB model. Previous work in finite temperature DFT used this
model at finite temperature, but always restricted to $\mathcal{N}=2$ \cite{SPB16}. Here we look at all $\mathcal{N}$, in order to accurately test PPLB.

Our first (and most important) result is already shown in Fig 1. For this simple model, the ansatz
behind PPLB is correct, and the PPLB yields the exact zero-temperature limit of the chemical potential.
From this fact (for any electronic system), all the subsequent deductions of PPLB follow.  It is comforting
to know this is true in the one case where $\mu$ can be found exactly.

\begin{figure}[htb]
  \includegraphics[width=\linewidth]{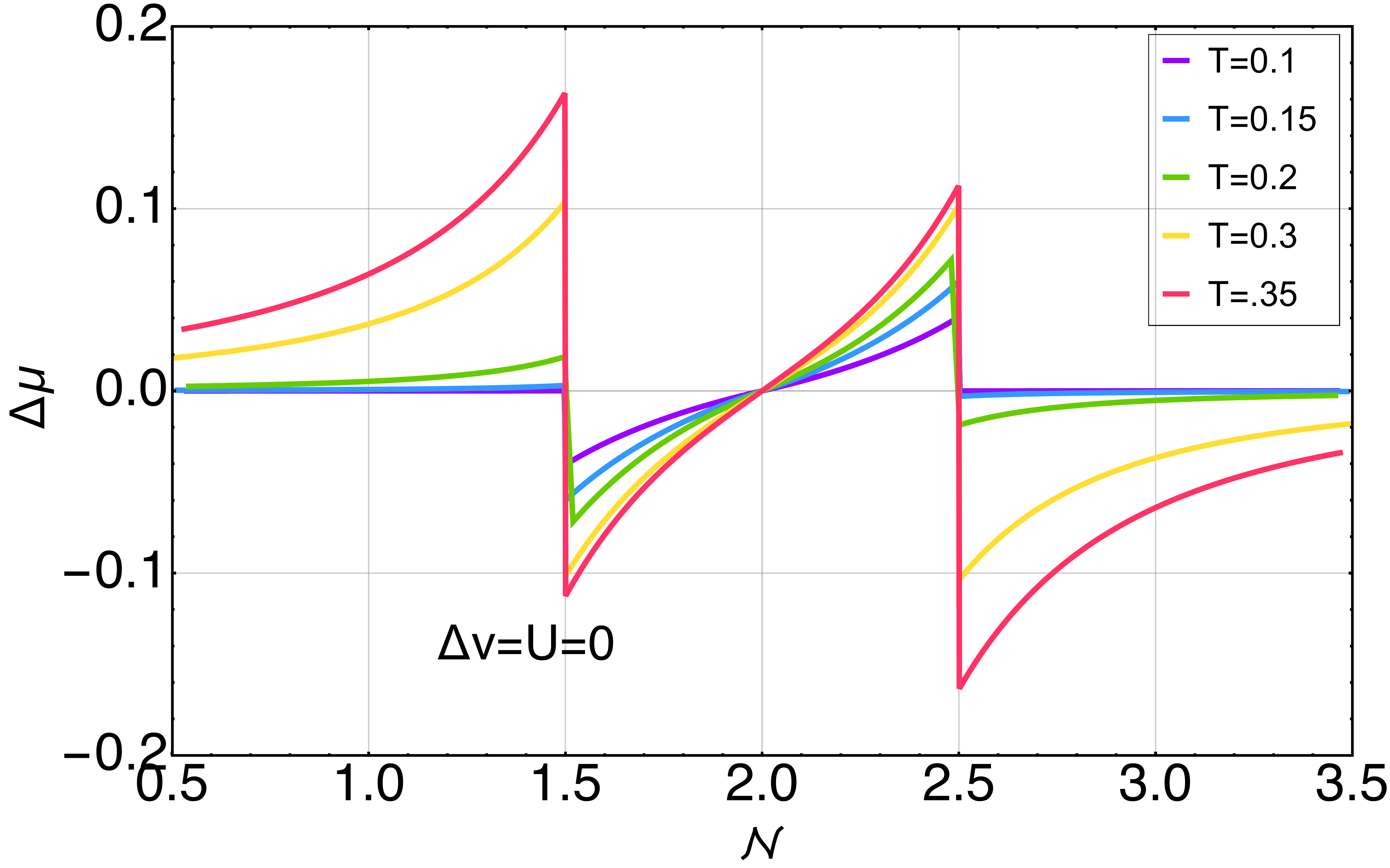}
  \caption{The absolute error in the approximate PPLB $\mu$ for the symmetric, tight binging case ($U=\Delta v= 0$), plotted with respect to the average particle number $\mathcal{N}$. Here $\Delta \mu=\mu-\mu^{^{PPLB}}$.}
  \label{plot2}
\end{figure}

But our next step is to explore PPLB for finite temperatures, and quantify how high in temperature it
can be considered to be working.  This is beyond the original intent of the model, which was designed
only to recover the zero-temperature limit. If we accept
errors in $\mu$ up to some threshold, say 0.1 $a.u.$, then the PPLB chemical potential works for $\Delta v = U=0$ until almost $\tau=0.3$, or about 100,000K for $2t=1$. This result is seen in Fig. 2. 

Next, in order to make this more relevant, we use the PPLB model to construct a PPLB approximation to the Helmholtz free energy, $A$. The exact gc partition function is 
\ben
Z(\tau, \mu)=\sum_{N,i}  g_{_{N}}^{(i)}  \exp{\bigg((\mu N-E^{(i)}_{_{N}})/ \tau \bigg)},  
\een

\noindent where $g_{_{N}}^{(i)}$ and $E^{(i)}_{_{N}}$ are the the degeneracy and energy of the $i$-th state for $N$ particles. The exact average particle number is then found via

\ben
\mathcal{N}(\tau,\mu) =-\tau \frac{d \log Z(\tau,\mu)}{d \mu},
\label{eq:N}
\een
\noindent so that the free energy can then be written as

\ben
A(\tau,\mathcal{N})= \mu(\tau, \mathcal{N}) \mathcal{N}+ \tau \log Z(\tau,\mu(\tau,\mathcal{N})),
\label{A}
\een
\noindent where $\mu(\tau, \mathcal{N})$ is the inverse of eq. \ref{eq:N}. As a step toward deriving eq. \ref{muPPLB}, we break down the derivation into two steps. First we introduce a simple (but different) continuous ground-state approximation, which includes only the ground states in the approximate partition function.  Such a partition function, denoted as $Z_{_{0}}$, is

\bea
Z_{_{0}}(\tau,\mu)=\sum_{N} g_{_{N}} \exp{\bigg((\mu N-E_{_{N}})/\tau \bigg)}.  
\eea

\begin{figure}[htb]
  \includegraphics[width=\linewidth]{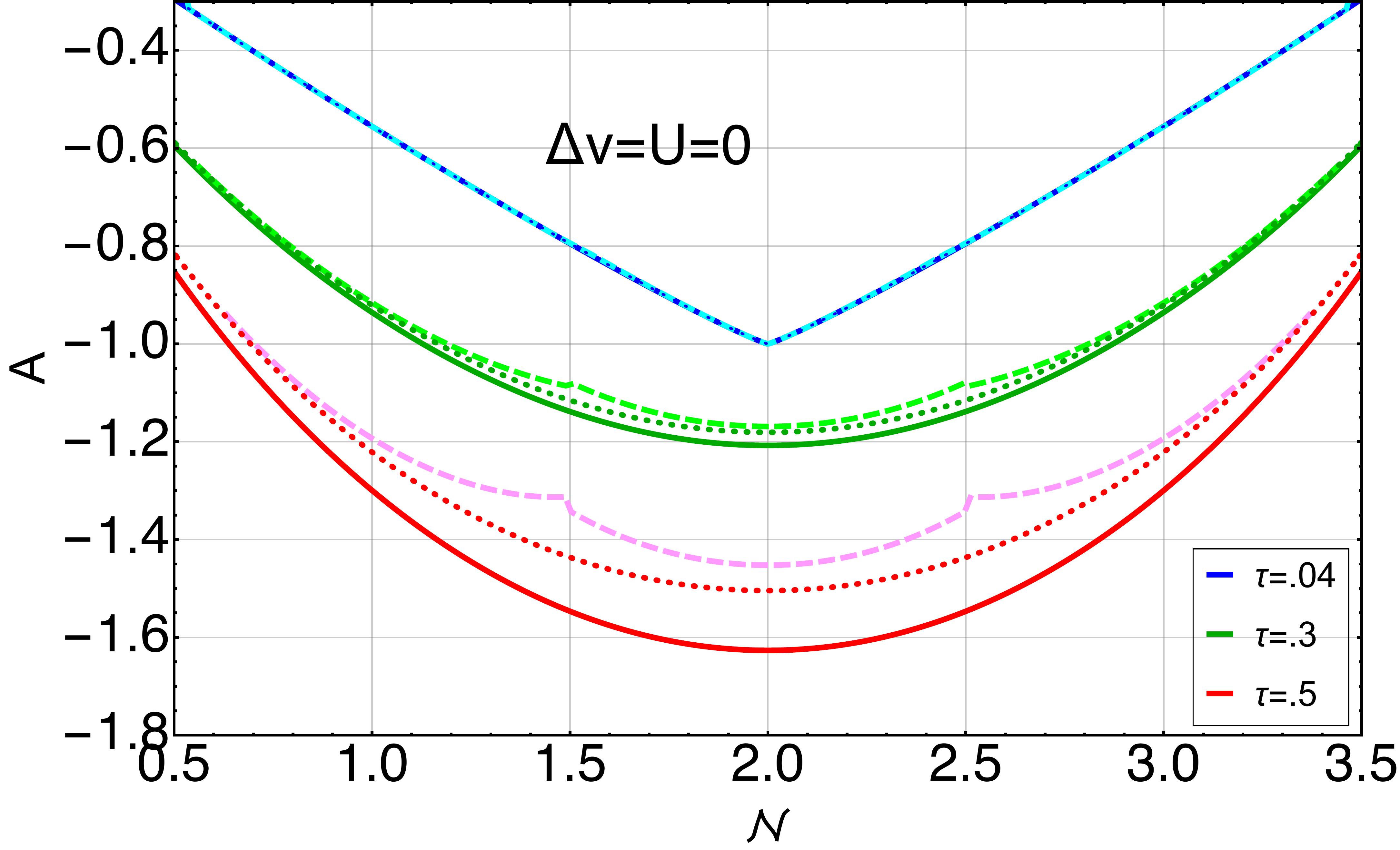}
  \caption{Free energy $A$, plotted for various temperatures, for $U=\Delta v=0$. Brighter dashed lined correspond to $A^{^{PPLB}}$, dotted lines are the ground state approximation to the free energy, $A_{_0}$, and solid lines are the exact values. }
  \label{plot3}
\end{figure}

\noindent For the Hubbard dimer at finite temperatures, this is a simple continuous function of $\mu$ with only five terms. Then,

\ben
A_{_{0}}(\tau,\mathcal{N})= \mu_{0}(\tau,\mathcal{N}) \mathcal{N}+ \tau \log Z_{_{0}}(\tau,\mu_{0}(\tau, \mathcal{N})),
\label{eqA0}
\een
\noindent where $\mathcal{N}_{0}(\tau,\mu)$ is found from plugging $Z_{0}$ into eq. \ref{eq:N}, and $\mu_{0}(\tau,\mathcal{N})$ is its inverse. This ground state approximation is plotted in Fig. \ref{plot3} as the dotted lines, and is a smooth well behaved function. $Z_{_{0}}$ is a better approximation than the PPLB, \textit{but} requires the ground state energies for all $N_{0}$ because $\mu^{^{PPLB}}$ is a piecewise function of $\mathcal{N}$, it is not found from a valid (or traditional) partition function. Instead, we define $A^{^{PPLB}}$ with eq. \ref{APPLB}, inserting eq. \ref{muPPLB} for $\mu_{0}$, and truncate $Z_{0}$ to the three nearest integers. 

\ben
A^{^{PPLB}}= \mu^{^{PPLB}} \mathcal{N}+\tau \log \widetilde{Z}^{^{PPLB}}(\tau,\mathcal{N},\mu^{^{PPLB}}),
\label{APPLB}
\een

\noindent where

\ben
\widetilde{Z}^{^{PPLB}}(\tau,\mathcal{N},\mu)=\sum_{J=M-1}^{M+1} g_{_{J}} \exp{\bigg((\mu J-E_{J})/\tau \bigg)},  
\label{eqZPPLB}
\een
\noindent and $M$ is the integer closest to $\mathcal{N}$. For the Hubbard dimer, this means that $\widetilde{Z}^{^{PPLB}}$ is a discontinuous, piecewise function.  While eq. \ref{eqZPPLB} is not a traditional partition function, as it is a function of $\mathcal{N}$, it still does rather well in approximating the free energy of the system. Notice that the difference between $A^{^{PPLB}}$ and $A_{_{0}}$ becomes negligible as $\tau \rightarrow 0$. 

A feature that makes the Hubbard dimer extremely useful in DFT studies is that one can make correlation
arbitrarily strong. For the symmetric case (that is at $\Delta v=0$), $U=2$ is the point at which it switches from weakly to strongly
correlated \cite{CFMB18,SB18}.
For strong asymmetry (when $\Delta v >> 1$), this happens near $U=\Delta v$ (see Fig 7 of Ref.\cite{CFMB18}).
In Fig \ref{plot3}, we show the performance of our PPLB free energy model when $U$ is small.  
In sharp contrast to semi-local approximations, it perfectly captures discontinuities at integer values, but 
artificially introduces steps at {\em half}-integers, which are noticeable when the value of $\tau$ is large enough.

While the symmetric case for 2 sites (and the homogeneous case for many sites) is the most frequently 
studied in many-body condensed matter physics, one must consider inhomogeneity to understand the density
functional aspects of the problem \cite{CFMB18}.  We next turn on significant asymmetry ($\Delta v=5$), and in Fig 4,
we plot PPLB for $U=2$ and $U=10$. Here a few things are noted. First, that using this $A^{^{PPLB}}$ gives a surprisingly accurate approximation to the free energy, even at finite temperatures. Second, that the largest magnitude of the absolute error \textit{always} appears at either $\mathcal{N}=2$ or $\mathcal{N}=$ half-integer, with the errors vanishing at the endpoints or when $\mathcal{N}=.5,3.5$, in this model system. This is in contrast to what was seen in Fig. \ref{mu}, where the errors in  $\mu^{^{PPLB}}$ \textit{always} vanish at $\mathcal{N} = 2$. Finally, the PPLB free energy works best for weak correlation ($\Delta v >> U$) and fails quantitatively for strong correlation ($\Delta v << U$), just as semilocal functionals do \cite{CFMB18,SB18}. Most importantly, as stated in the original PPLB paper, the approximation becomes exact at the zero temperature limit, capturing the derivative discontinuities.

\begin{figure}[htb]
  \includegraphics[width=\linewidth]{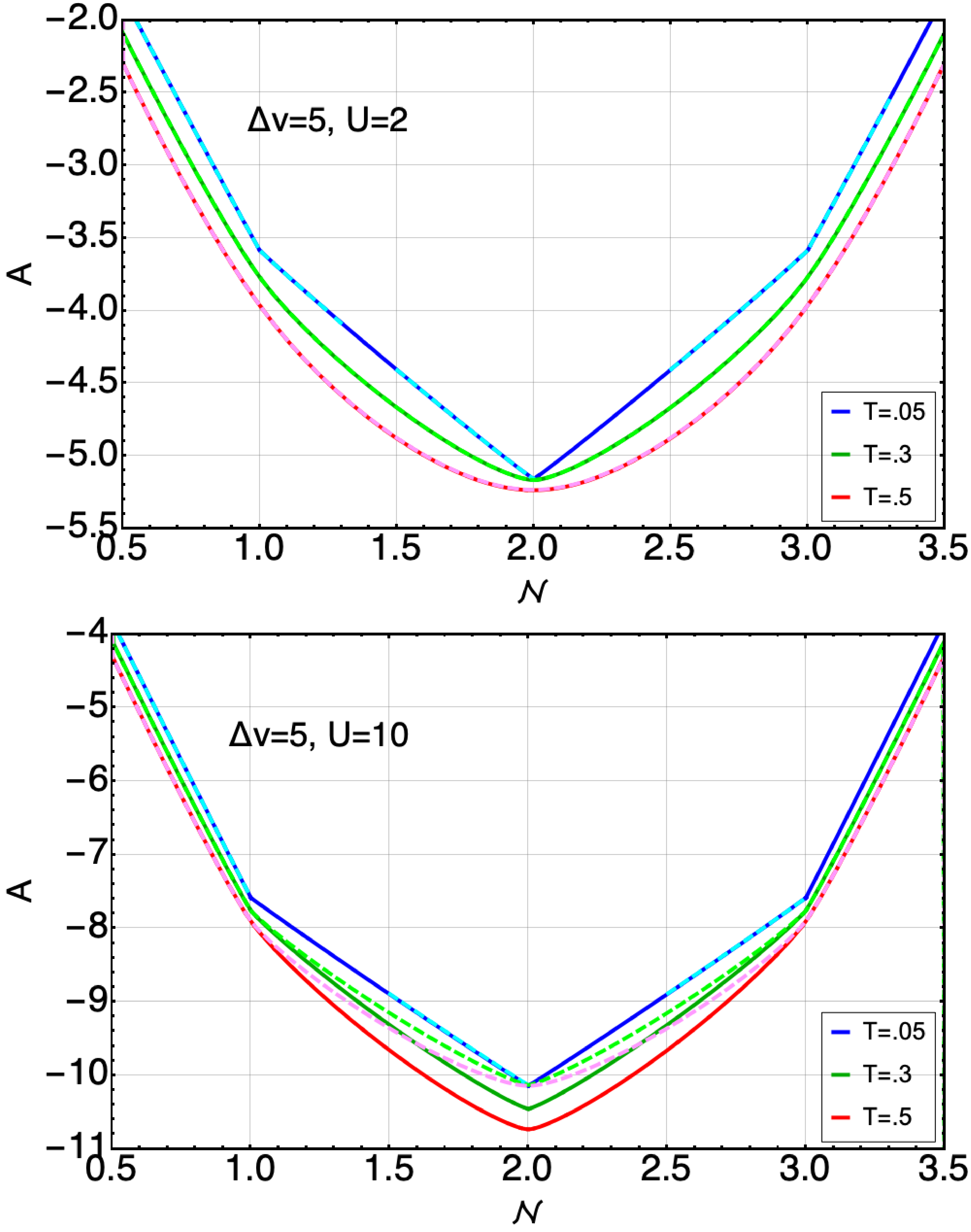}
  \caption{Free energy, for weakly correlated, (top panel), and strongly correlated regimes (bottom panel), plotted with respect to the average particle number $\mathcal{N}$.}
  \label{plot4}
\end{figure}

\begin{figure}[htb]
  \includegraphics[width=\linewidth]{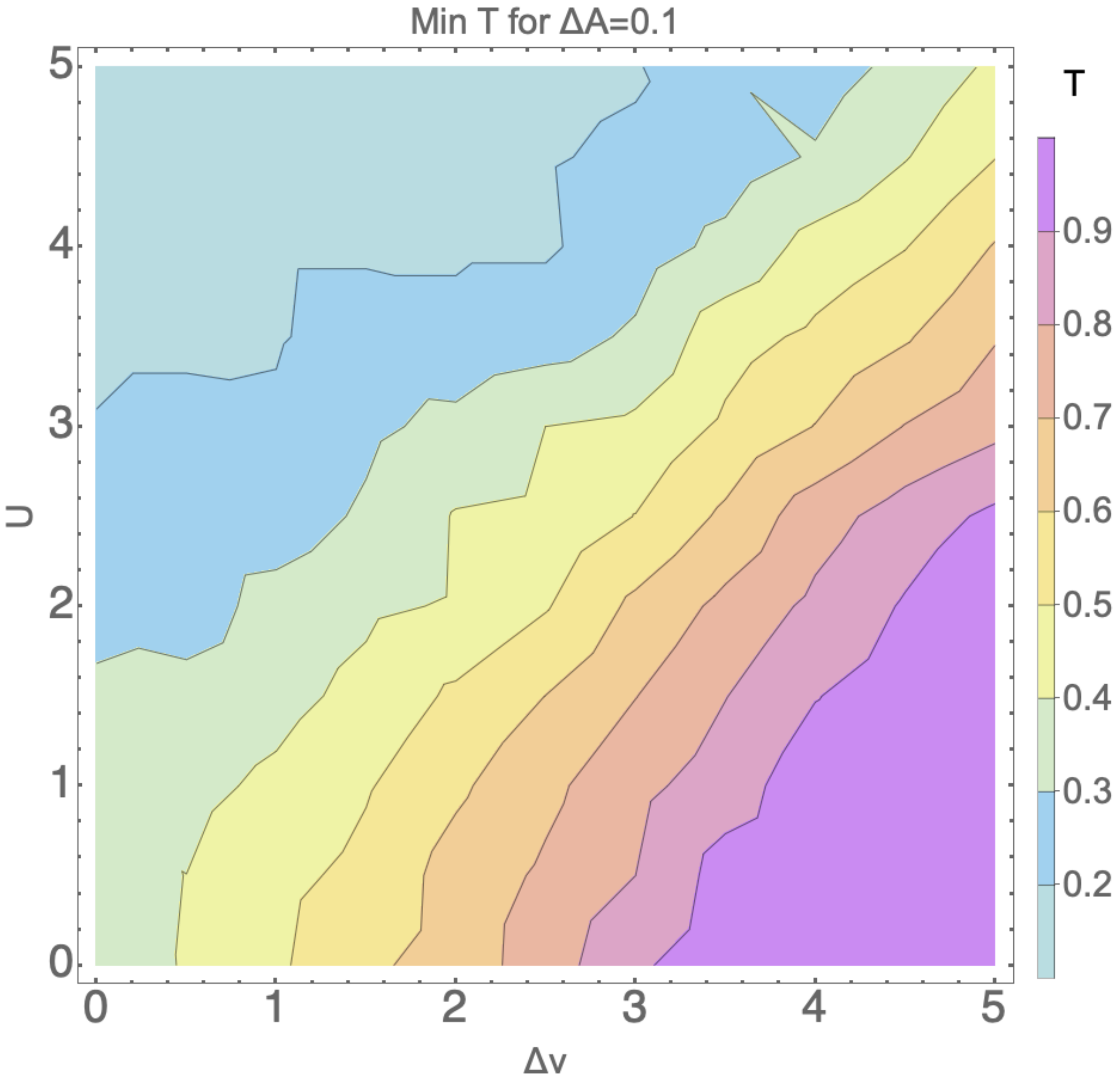}
    \caption{Contour plot of the minimum value of the temperature $\tau$  which gives absolute errors to the free energy $\Delta A=A-A^{^{PPLB}}$ of $0.1$, plotted for various $\Delta v$ and $U$.}
  \label{plot5}
\end{figure}

We can summarize the efficacy of the PPLB absolute error in the free energy, $\Delta A$, with a contour plot in the $(U-\Delta v)$ plane. We make a crude contour plot of the temperatures at which the absolute error in the free energy using this PPLB approximation is no greater that 0.1 $a.u$. for any value of $\mathcal{N}$, and the colors of the contour correspond to the temperature at which $\Delta A = 0.1$ $a.u$ occurs.  Fig. \ref{plot5} shows these results. This calculation uses a coarse grid due to computational cost, caused by the discontinuous changes in the PPLB errors,
but the structure is clear. There is an obvious divide seen between the strongly and weakly correlated regimes \cite{CFMB18,SB18}. Clearly, the PPLB approximation works better for the weakly correlated regime where steps are small, and $\tau$ reaches high temperatures before the error reaches $0.1$ $a.u.$ Likewise, as the value of $U$ increases to a point $\Delta v<<U$, then the maximum temperature for our benchmark error decreases substantially.  

Lastly, we consider how one might extend the temperature range of the accuracy of the PPLB free energy. We simply include the most relevant terms beyond those included in $\mu^{^{PPLB}}$. Since the PPLB partition function includes only 
the ground-state contribution for each $N$, the addition to include the first excited
state energy to all $N$ seems to be the most obvious. This correction can be included in eq. \ref{muPPLB}, by simply replacing each $g_{_{M}}$ with $\widetilde{g}_{_{M}}$, where 



\ben
\widetilde{g}_{_{M}}= g_{_{M}} + g_{_{M}}^{(1)}\exp{(-(E_{_M}^{(1)}-E{_{_M}})/\tau)},
\label{eq.g*}
\een

\noindent $g_{_{M}}^{(1)}$ and $E_{_M}^{(1)}$ correspond to the degeneracy and first excited state for $M$ particles.

\begin{figure}[htb]
  \includegraphics[width=\linewidth]{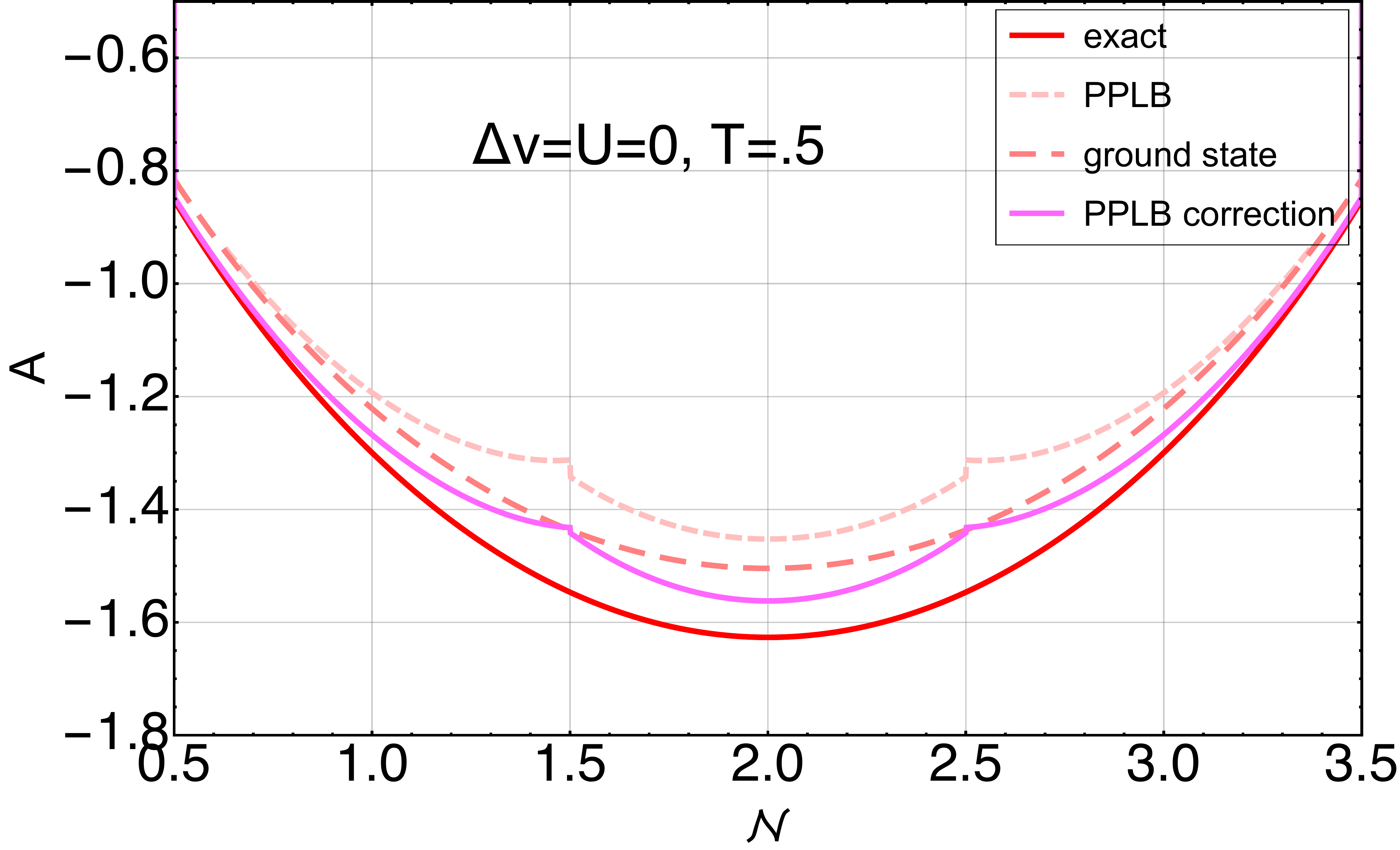} 
   \caption{Comparing $A$ (eq. \ref{A}), $A^{^{PPLB}}$ (eq. \ref{APPLB}), $A_{0}$(eq. \ref{eqA0}), and $A^{*}$(eq. \ref{eq.g*}),  for $\tau=.5$. }
  \label{plot7}
\end{figure}

\begin{figure}[htb]
  \includegraphics[width=\linewidth]{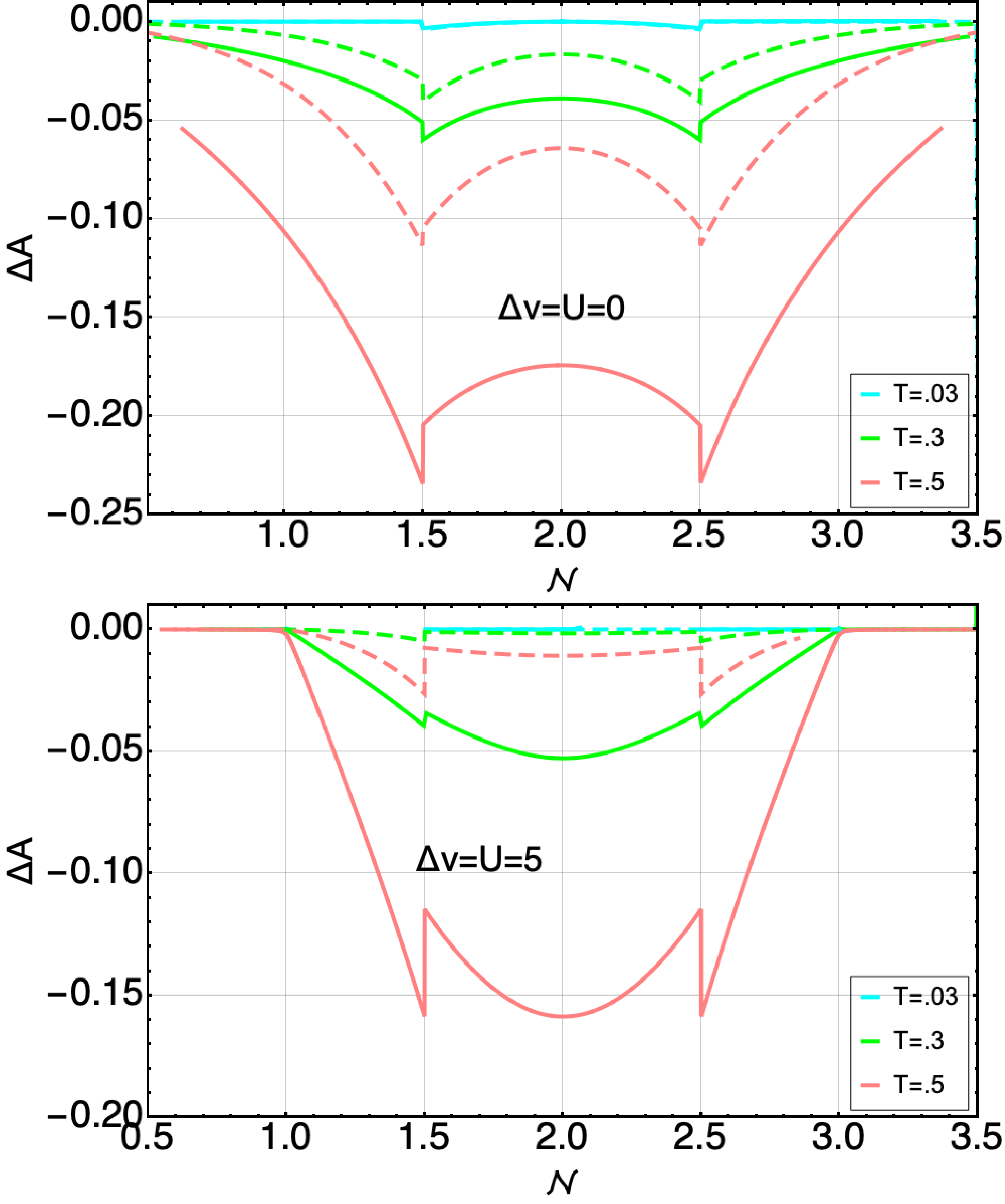}
  \caption{Comparing the absolute errors in the energy, $\Delta A$. Solid lines are eq. \ref{APPLB} and dashed lines are eq. \ref{eq.g*} plugged into eq. \ref{muPPLB}. }
  \label{plot8}
\end{figure}

In Fig. \ref{plot7} we compare the exact free energy, the PPLB approximation, the ground state approximation, and our correction to the PPLB free energy. Clearly, there is an improvement when compared to the PPLB formalism at the largest quantitative errors. When the half-integers are the points of largest quantitative errors, that is when $\Delta v >> U$ (and thus weakly correlated), this would cause havoc for any
derivatives of the energy in a real system, such as those used to find densities. These steps are places where one value
of a parameter is suddenly swapped with another and when the nearest integer changes. So these parameters include
the fundamental gap, the ionization energy, and the degeneracies of the energy levels. Any simple smoothing
function, or correction to the PPLB could eliminate these. 
In Fig. \ref{plot8}, we compare the results for $\Delta A$ with the PPLB approximation, and our correction for various $\tau$. In these figures, it is clearly seen that our correction provides a substantial improvement to $A^{^{PPLB}}$, even at higher values of $\tau$.

To put this work in context, our Hubbard dimer
looks nothing like the systems used in KS DFT warm dense matter simulations \cite{KDD08,KHBH01,D03,NHKF08}.  However, such calculations often have
features driven by the underlying molecular structure, for which energetic consequences of the derivative
discontinuity are known to be quantitatively relevant. Our study here has focussed on the full chemical potential and free energy of the system, not the exchange-correlation
contributions that are so important in density functional theory.  Our general results apply to finite temperature simulations of
localized electrons in any formalism, and so can be used to gain insight into WDM simulations of any kind. The relevance
 of the PPLB reasoning, and its extension to free energies at finite temperatures given here, is likely unknown in the general WDM community.
For example, ionization lowering \cite{JLHS20,SP66} can now be related to the behavior of both the chemical potential
and the free energy. Our work is in the spirit of simple conditions
at zero-temperature \cite{SPB16}.  We have found that (1) the PPLB was correct as derived for the limit $\tau \rightarrow 0 $. We have shown what an exact treatment should do, and how well the PPLB model 
captures this, and (2) how one can understand up to what temperatures it will be accurate. We have (3) used the PPLB formalism to accurately simulate the free energy at finite temperatures, even though this approximation to the chemical potential was originally intended for the zero temperature limit. Last, we have (4) provided a correction to this PPLB model to make it even more useful. 

In terms of real-world applications, for any
finite temperature KS DFT calculation of a molecular system \cite{GCG10}, one could easily construct the
PPLB free energy, using only total energy differences as inputs. These could come from either a highly
accurate quantum chemical calculation, or even a DFT calculation. The error estimates requiring excitation
energies could be extracted from TDDFT \cite{CFMB18,M16} or an ensemble DFT calculation \cite{YPBU17,SB18}. Then an accurate picture of the free energy can be calculated
up to reasonable temperatures using PPLB.

\sec{Acknowledgements} F.S and K.B acknowledge support from the
Department of Energy, Award No. DE-FG02-08ER46496.
\clearpage
\bibliography{perdew_ref}
\label{page:end}
\end{document}